\documentclass[10pt, preprintnumbers,amsmath,amssymb]{revtex4}

\usepackage{latexsym}
\usepackage{amssymb}
\usepackage{epsfig,amsmath,graphics}
\usepackage{epstopdf}
\usepackage{comment}

\newcommand{\GeV}{\text{GeV}}
\newcommand{\MeV}{\text{MeV}}
\newcommand{\cm}{\text{cm}}
\newcommand{\Hz}{\text{Hz}}

\newcommand{\shot}{\text{shot}}
\newcommand{\static}{\text{static}}
\newcommand{\eV}{\text{eV}}

\newcommand{\Bext}{B_\text{ext}}
\newcommand{\vBext}{\vec{B}_\text{ext}}

\begin{document}

\title{New Observables for Direct Detection of Axion Dark Matter}

\author{Peter W. Graham}
\affiliation{Stanford Institute for Theoretical Physics, Department of Physics, Stanford University, Stanford, CA 94305}

\author{Surjeet Rajendran}
\affiliation{Stanford Institute for Theoretical Physics, Department of Physics, Stanford University, Stanford, CA 94305}

\begin{abstract}
We propose new signals for the direct detection of ultralight dark matter such as the axion.  Axion  or axion like particle (ALP) dark matter may be thought of as a background, classical field.  We consider couplings for this field which give rise to observable effects including a nuclear electric dipole moment, and axial nucleon and electron moments.  These moments oscillate rapidly with frequencies accessible in the laboratory, $\sim$ kHz to GHz, given by the dark matter mass. Thus, in contrast to WIMP detection, instead of searching for the hard scattering of a single dark matter particle, we are searching for the coherent effects of the entire classical dark matter field. We calculate current bounds on such time varying moments and consider a technique utilizing NMR methods to search for the induced spin precession. 
The parameter space probed by these techniques is well beyond current astrophysical limits and significantly extends laboratory probes.  Spin precession is one way to search for these ultralight particles, but there may well be many new types of experiments that can search for dark matter using such time-varying moments.
\end{abstract}

\maketitle

\tableofcontents

\section{Introduction}
The existence of dark matter is concrete evidence for physics beyond the standard model. It is reasonable to expect the dark matter to interact non-gravitationally with the standard model. The identification of such interactions may allow new probes of the Universe and unveil new structures in particle physics.  The generic expectation of the existence of new physics at the weak scale lead to the Weakly Interacting Massive Particle (WIMP) hypothesis - the dark matter is a weak scale particle that interacts with the standard model with weak scale cross-sections.  A variety of experimental approaches have been developed to test the WIMP hypothesis. These include techniques to observe the direct scattering of WIMP particles with nuclei \cite{Goodman:1984dc} and electrons  \cite{Graham:2012su, Essig:2011nj}, the detection of cosmic rays produced from the annihilation \cite{Rudaz:1987ry,  Galli:2009zc, Slatyer:2009yq} or decay   \cite{Arvanitaki:2008hq, Arvanitaki:2009yb} of dark matter particles as well as searches in colliders for weakly interacting particles   \cite{Sigurdson:2004zp, Rajaraman:2011wf, Bai:2010hh}. These techniques have been deployed in a variety of dedicated experiments that have placed significant constraints on the parameter space of viable WIMP dark matter   \cite{Ahmed:2011gh, :2013doa, Meade:2009iu, Essig:2012yx}.  Further, direct probes at the Large Hadron Collider (LHC) of frameworks such as supersymmetry that have provided the theoretical support for WIMP dark matter have placed stringent constraints on such models \cite{Lowette:2012uh}. In fact, these stringent constraints arise from the assumption that the new physics at the LHC  always produces a metastable WIMP particle. Indeed, the bounds on these frameworks are significantly alleviated by allowing for such WIMP particles to decay rapidly within the collider \cite{Graham:2012th},  precluding a cosmological role for them. 

Given our ignorance of the ultra-violet framework of particle physics and the nature of dark matter, it is important to develop techniques to search for a wide variety of interactions that could be carried by the dark matter. Ultra-light scalars such as axions ($a$) and axion-like-particles (ALPs) with masses $m_a$ significantly smaller than the weak scale are a well motivated class of dark matter candidates. These particles emerge naturally as the Goldstone bosons of global symmetries that are broken at some high scale $f_a$ \cite{Peccei:1977hh, Peccei:1977ur, Weinberg:1977ma, Wilczek:1977pj, Kim:1979if, Shifman:1979if, Dine:1981rt,  Zhitnitsky:1980he, Khlopov:1999rs} (see section \ref{Sec:Overview} for an overview of such particles). Their Goldstone nature is manifest in their derivative interactions with the standard model:

\begin{equation}
\label{Eqn:Terms}
\frac{a}{f_a} F_{\mu \nu} \tilde{F}^{\mu \nu}, \, \frac{a}{f_a} G_{\mu \nu} \tilde{G}^{\mu \nu}, \frac{\partial_{\mu} a}{f_a} \bar{\Psi}_{f} \gamma^{\mu} \gamma_5 \Psi_{f}.
\end{equation}
Here, $F_{\mu \nu}$ and $G_{\mu \nu}$ represent the field strengths of electromagnetism and QCD respectively and $\Psi_f$ denotes a standard model fermion. The first of these interactions couples axions (and ALPs) to photons and is used in a variety of experiments to search for the axion. These include methods to search for the conversion of dark matter axions into photons in the presence of a background magnetic field \cite{Sikivie:1985yu, Asztalos:2009yp}, the detection of axions produced in the Sun \cite{Irastorza:2006gs} and axion-aided transport of photons through optical barriers \cite{Ehret:2009sq}. These experiments can search for axions with $f_a \lessapprox 10^{12}$ GeV, with limited ability to go above this scale. 

In this paper, we argue that the other operators in \eqref{Eqn:Terms} can also be used to search for axion (and ALP) dark matter. These operators are particularly useful in probing dark matter axion parameter space where $f_a \gtrapprox 10^{14}$ GeV.  An ultra-light particle like the axion ($m_a \ll $ meV) can be a significant fraction of the dark matter only if it has a large number density,  leading to large field occupation numbers. Consequently, axion dark matter can be viewed as a background classical field oscillating at a frequency equal to its mass  \cite{Dine:1982ah, Preskill:1982cy}. Conventional axion experiments search for energy deposition from this background classical field. But, a background classical field can also lead to additional physical effects such as giving rise to phase differences in local experiments. For example, gravitational wave experiments   
  \cite{Abbott:2007kv, Accadia:2012zz, Dimopoulos:2007cj, Dimopoulos:2008sv, Hogan:2010fz, Graham:2012sy} aim to detect gravitational waves through the phase differences created by the wave instead of the unobservably small rate with which a single graviton would scatter and deposit energy in a detector. Similarly,  we show that the classical axion (and ALP) dark matter field through its interaction with the operators in \eqref{Eqn:Terms}, leads to energy (and phase) differences in atomic systems. These phase differences manifest themselves as time varying moments that can be used to search for such dark matter.  Essentially, axion dark matter can be thought of as an oscillating value of the Strong CP angle $\theta_\text{QCD}$, which can lead to new ways to detect it. 
  
For example, it was pointed out in \cite{Graham:2011qk} that the second operator in \eqref{Eqn:Terms} gives rise to a time varying nuclear electric dipole moment. These moments oscillate at a frequency equal to the mass of the axion, which can span the frequency space kHz-GHz. Even though the axion arises from ultra-high energy physics, its mass is small enough to be accessible in the laboratory. These frequencies though are rapid enough that conventional laboratory searches for such moments (such as a nuclear electric dipole moment) would have a reduced sensitivity to them since such experiments (for example, \cite{Romalis:2001qb}) gain sensitivity through long ($\gtrapprox 1$ s) interrogation times. Further, conventional searches for time dependent moments have focussed on time variations occurring over the Hubble scale \cite{Hubble}. These time variations require the existence of scalar fields with masses comparable to the Hubble scale and such supremely light masses are difficult to obtain without fine-tuning \cite{Dvali:2001dd}. In contrast, time variations in the frequency range kHz - GHz emerge naturally in many axion (and ALP) models (see section \ref{Sec:Overview}). 

In this paper, we will show that the time varying moments induced by the dark matter axion (and ALP) can couple to nuclear or electronic spin leading to a precession of the spin. While there may be other experimental strategies to measure these time varying moments, we highlight the technique described in \cite{NMR paper}. In this technique, the induced spin precession changes the magnetization of a sample of material, which  can be observed with precision magnetometry. The signal in such an experiment benefits from the large number of spins that can be obtained in a condensed matter system and the availability of high precision SQUID and SERF magnetometers. The time varying nature of the signal can be used to devise resonant schemes that can significantly boost their detectability. Further, since this time variation occurs at a frequency set by fundamental physics, it can also be helpful in combatting systematic noise sources that are encountered in searching for a time independent moment. 

We begin by briefly reviewing the physics of the axion in section \ref{Sec:Overview}. In this section, we discuss the theoretical origins of axions and ALPs, their parameter space, the current constraints on this parameter space and the region where they can be dark matter. We also review the salient features of axion (and ALP) dark matter. Following this review, we discuss current bounds on such moments and estimate the reach of the precision magnetometry experiments discussed above to probe this parameter space.

\section{Axions and Axion-like-Particles (ALPs)}
\label{Sec:Overview}

Axions and ALPs are generically expected in many models of physics beyond the standard model \cite{Svrcek:2006yi}. They are the Goldstone bosons of global symmetries that are broken at some scale $f_a$. The Goldstone nature of their origin is manifest in the fact that all their interactions are suppressed by the scale $f_a$ and they are coupled derivatively in these interactions (for example, the operators in \eqref{Eqn:Terms}). If they were pure Goldstone bosons, they would be completely massless and would not be the dark matter. However, their pure Goldstone nature could be broken leading to a generation of a mass for them. 

This mass generation can occur if the associated global symmetry is anomalous. For example, when this global symmetry has an anomaly with QCD, non-perturbative dynamics at the QCD scale can generate a mass $m_a \sim \frac{\Lambda_{\text{QCD}}^2}{f_a}$ or more precisely
\begin{equation}
\label{eqn: axion mass}
m_a \approx 6 \times 10^{-10} \text{ eV} \, \left(\frac{10^{16} \text{ GeV}}{f_a}\right)
\end{equation}
for the Goldstone boson with about a 10\% uncertainty due to QCD \cite{Cadamuro:2010cz}. This Goldstone boson is called the axion. While it was initially introduced to dynamically solve the strong CP problem \cite{Peccei:1977hh, Peccei:1977ur}, the underlying mechanism responsible for its dynamics is very general. For example, if the associated symmetry had a mixed anomaly with some other gauge group that also had strong dynamics, that Goldstone boson would also acquire a mass $\sim \frac{\Lambda_s^2}{f_a}$, where $\Lambda_s$ is the scale where the new gauge group becomes strong. A small breaking of the global symmetry is another source of mass.  One source of such a small breaking could be quantum gravity which is generically expected to violate global symmetries. In this case, the Goldstone bosons would acquire a mass proportional to the breaking of this global symmetry, leading to a tiny mass for them. Goldstone bosons that do not acquire a mass from QCD are called axion-like-particles (ALPs). Thus, while the mass of the axion depends upon only one free parameter $f_a$, the mass of ALPs depends upon additional parameters. 

In this paper, we will be interested in ALPs whose masses are comparable to that of the axion ($\sim$ kHz - GHz). The experimental limits and signatures discussed will apply to both axions and ALPs and henceforth we will refer to them both as ALPs. There  are several astrophysical bounds on the ALP operators in \eqref{Eqn:Terms} and they rule out ALPs with $f_a \lessapprox 10^{9}$ GeV \cite{Raffelt:2006cw}. Theoretical prejudice suggests that $f_a$ should lie around the fundamental scales of particle physics such as the grand-unified ($\sim 10^{16}$ GeV) or the Planck ($\sim 10^{19}$ GeV) scales, where we expect other symmetries of nature to be broken \cite{Svrcek:2006yi}. A weak upper bound of  $f_a \lessapprox 10^{12}$ GeV can also be placed on the axion based upon its cosmological abundance \cite{Dine:1982ah, Preskill:1982cy}. This bound is specific to the QCD axion and arises by assuming that the axion field is displaced $\mathcal{O}\left(1\right)$ from its minimum in the early Universe. This bound can be relaxed, allowing for larger values of $f_a$,  if the field displacement was not as large. In fact, the maximum field displacement scales  $ \propto \frac{1}{\sqrt{f_a}}$ implying that $f_a \approxeq 10^{16}$ GeV (the grand unified scale) would be allowed as long as the initial displacement was $\sim \mathcal{O}\left(1 \% \right)$.

As argued in \cite{Linde:1987bx}, inflationary cosmology can provide a natural mechanism that would allow a range of initial values for the axion field. As long as the scale of inflation is lower than $f_a$, the pre-inflationary space-time can have a generic inhomogeneous distribution of the axion field. Inflation can make any small part of this initial space-time into our Hubble patch, allowing for a uniform axion field value throughout our patch. The field value in our Hubble patch would be equal to the local field value of the pre-inflationary space-time that inflated to become our patch. Since a range of initial field values are scanned in the pre-inflationary space-time, the axion field can take any value in our patch. It is difficult to estimate the likelihood of any particular value in our  patch since inflationary space-times do not possess a natural measure \cite{Linde:2007nm}. 

For any $f_a \gtrapprox 10^{9}$ GeV, a cosmologically viable axion field exists as long as the initial field value is appropriately small ($\propto \frac{1}{\sqrt{f_a}}$). In the canonical axion window $f_a \sim 10^{12}$ GeV, it is assumed that it  is natural for the axion field to have a large  $\mathcal{O}\left(1\right)$  initial displacement, forcing the scale $f_a$ to be much lighter than the fundamental scales of particle physics. But, having $f_a$ much lighter than these fundamental scales creates another hierarchy problem. The constraint  \cite{Dine:1982ah, Preskill:1982cy} on the cosmological abundance of the axion requires one small number - this could be the initial value of the axion field or the scale $f_a$ (measured in terms of the fundamental scales) or any combination of these. Given our ignorance of the ultraviolet structures of particle physics and the difficulties of obtaining well defined measures on initial conditions in inflationary cosmology, there is no strong reason to prefer any particular value of $f_a$. 

It should be noted that the details of the cosmological bound discussed above are specific to the QCD axion since this bound depends upon the details of the mass generation mechanism. For a general ALP, depending upon the mass generation mechanism, similar bounds could be placed. However, much like the case of the axion, bounds based on the cosmological abundance of ALPs are subject to similar uncertainties. 

It is thus important to search for ALPs over their entire range of parameter space, just above $f_a \gtrapprox 10^{9}$ GeV. The axion can constitute a significant fraction of the dark matter when $f_a \gtrapprox 10^{11}$ GeV, while for a generic ALP, the exact region where it can be the dark matter depends upon its mass generation mechanism. The phenomenology of cosmic ALP dark matter is insensitive to the details of the mechanism responsible for generating its mass. 

The classical field that describes ALP dark matter can be expressed as  $a_0 \cos \left(m_a t\right)$. The amplitude $a_0$  is obtained by setting the energy density in the field $\frac{1}{2} m_a^2 a_0^2$ equal to the local dark matter density $\rho_{\text{dm}}\sim 0.3 \, \frac{\text{GeV}}{\text{cm}^3}$. This energy density can be understood as the result of the oscillations of the classical dark matter ALP field with amplitude $a_0$ and frequency $m_a$. The temporal coherence of these oscillations in an experiment is limited by  motion through the spatial gradients of the ALP field. The gradients are set by the de Broglie wavelength of the ALP $\frac{1}{m_a v} \sim  1000 \, \text{km} \, \left(\frac{\text{MHz}}{m_a}\right)$, where $v \sim 10^{-3}$ is the galactic virial velocity of the ALP dark matter. Since the velocity between the  experiment and the dark matter is also  $v$, the time $\tau_a$ over which the ALP will interact coherently is at least $\tau_a \sim \frac{2 \pi}{m_a v^2} \sim 10^6 \, \frac{2 \pi}{m_a} \sim  1 \,s \, \left(\frac{\text{MHz}}{m_a}\right)$. In other words, the ALP's frequency $m_a$ is broadened by its  kinetic energy $m_a v^2$. Lighter ALPs, corresponding to larger values of $f_a$, are coherent for longer. For the rest of this paper, we will assume that ALPs constitute a significant fraction of dark matter and propose techniques utilizing their classical nature to search for them.  For another proposal using the fact that the axion acts like a classical field causing an oscillating $\theta$ angle, but using a different axion coupling from us, see for example \cite{Hong:1991fp}.

\section{Current Searches}
\label{sec: current searches}

\begin{figure}
\begin{center}
\includegraphics[width=5 in]{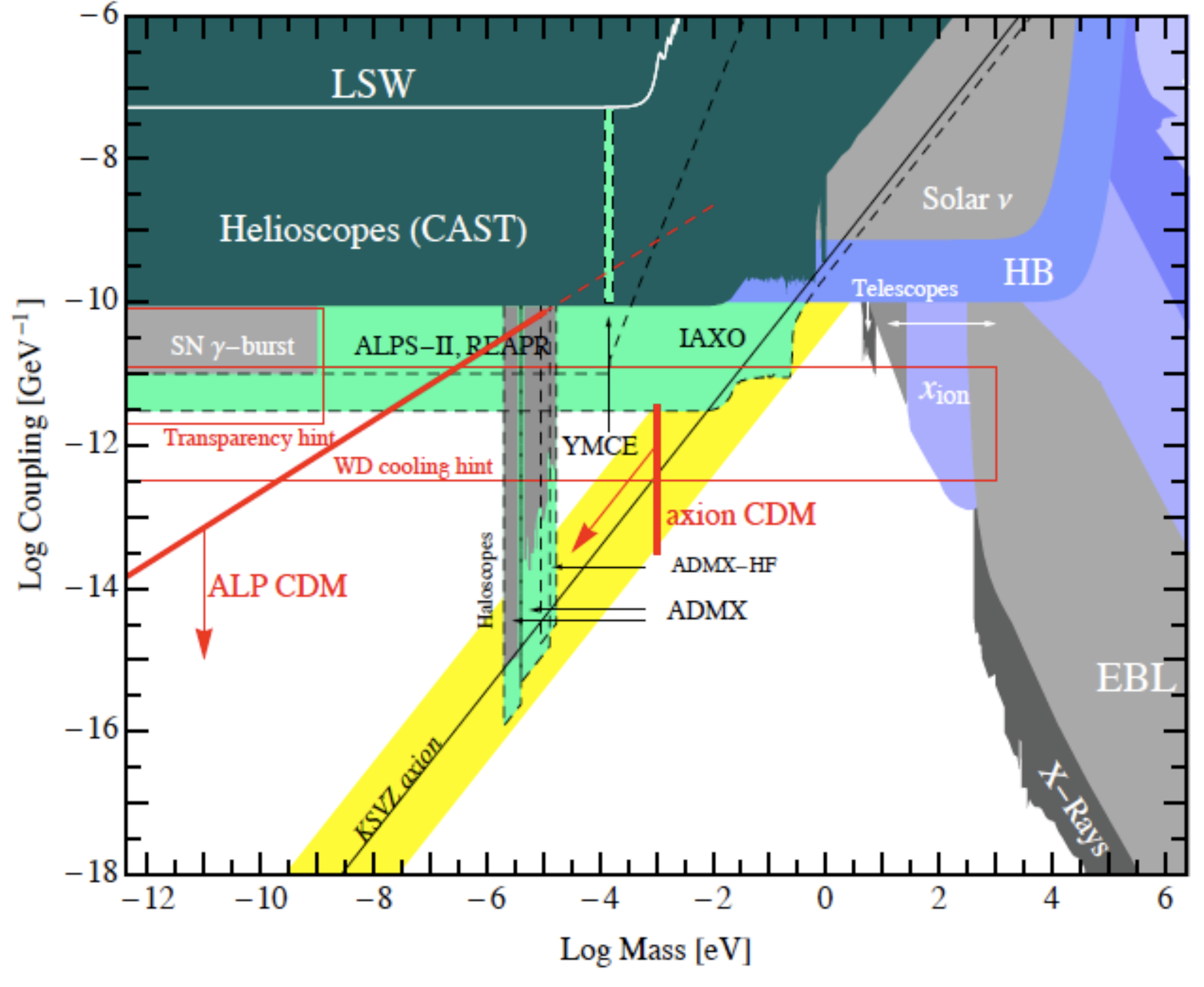}
\caption{ \label{Fig:photon}  Reproduced with permission from Fig.~2 of A.~Ringwald \cite{Ringwald:2012hr}, this figure is adapted from  \cite{Hewett:2012ns, Arias:2012az, Cadamuro:2011fd} (see also \cite{Ringwald:2012cu}).  ALP parameter space in axion-photon coupling (as in Eq.~\eqref{eqn: axion photon coupling}) vs mass of ALP.  The QCD axion is the yellow band.  The width of the yellow band gives an indication of the model-dependence in this coupling, though the coupling can even be tuned to zero.}
\end{center}
\end{figure}

Essentially all experiments attempting to directly detect axions or ALPs use the ALP coupling to electromagnetic fields
\begin{equation}
\label{eqn: axion photon coupling}
\mathcal{L} \ni g_{a\gamma\gamma} a F \tilde{F}
\end{equation}
where $F_{\mu \nu}$ is the field strength of electromagnetism and $a$ is the ALP field.
Such fields are easily manipulatable in the laboratory.  This has allowed a wide range of tests from microwave cavity experiments such as ADMX, to helioscopes such as CAST, to light-through-walls experiments such as ALPS \cite{Asztalos:2009yp, Irastorza:2006gs, Ehret:2009sq}.  There are also many astrophysical observations which limit this coupling of axions and ALPs to photons \cite{Raffelt:2006cw}.  A summary of all current constraints on this parameter space is reproduced in Fig.~\ref{Fig:photon} from \cite{Ringwald:2012cu} (see also \cite{Hewett:2012ns, Arias:2012az}).  Although this space is well-covered by experiments and astrophysical bounds at higher mass and coupling $g_{a\gamma\gamma}$, it is challenging to search in the low mass and coupling region.  A large piece of the parameter space for light ALPs is currently not reachable.  Indeed, most of the masses for which the QCD axion could be dark matter are not reachable by current experiments.

For the QCD axion there is a constrained set of predictions for $g_{a\gamma\gamma}$ as a function of the axion decay constant $f_a$, $g_{a\gamma\gamma} \propto f_a^{-1}$.  Thus the QCD likely lies in the yellow band in Fig.~\ref{Fig:photon}.
It may be the dark matter over a wide range of masses $\text{meV} \lesssim m_a \lesssim 10^{-12}$ eV, with axion decay constant anywhere in the range $f_a \gtrsim 10^{10}$ GeV all the way up to the Planck scale $\sim 10^{19}$ GeV.  The only experiment which can currently reach the QCD axion in this range is ADMX and it cannot probe the region of high $f_a \gtrsim 10^{13}$ GeV, or $m_a \lesssim 10^{-6}$ eV.  Coming upgrades to ADMX, such as ADMX-HF, may probe higher masses (lower $f_a$).  However, it is very challenging for microwave cavity searches to get to higher $f_a$ because these experiments search for axion to photon conversion through the coupling in Eq.~\eqref{eqn: axion photon coupling}.  The amplitude for this process necessarily goes as the square of the coupling $\propto f_a^2$.  Further, the cavity must be on resonance with the axion mass (frequency) in order to enhance the signal.  This requires the cavity to be approximately the size of the axion wavelength $\sim m_a^{-1}$.  For GUT scale axions $f_a \sim 10^{16}$ GeV gives $m_a^{-1} \sim 300$ m, which makes for a rather large cavity.  If the cavity size cannot be increased with the wavelength, then the sensitivity of the experiment will fall off even more rapidly with increasing $f_a$.  While microwave cavities make excellent axion detectors for the lower $f_a$, they are many orders of magnitude away from detecting axions with higher $f_a$.  Similarly, other proposals using the coupling to electromagnetism in Eq.~\eqref{eqn: axion photon coupling}, e.g.~the interesting, recent proposal of using a dish detector \cite{Horns:2012jf}, may work at lower $f_a \lesssim 10^{13}$ GeV, but cannot reach higher $f_a$.
 
Given how well-motivated axion dark matter is, it is important not to miss a such a large piece of its parameter space.  It is therefore crucial to design experiments that can detect axions or ALPs with masses below $\mu$eV.  This is clearly challenging using the coupling in Eq.~\eqref{eqn: axion photon coupling}.

\section{Axion-EDM Coupling}
\label{Sec: EDM}



There are two general problems to using the axion-photon coupling Eq.~\eqref{eqn: axion photon coupling} for detection of light, weakly-coupled axions.
First, all experiments are measuring rates for axion to photon conversion so they go as amplitude squared $\propto g_{a\gamma\gamma}^2$.  In fact, in the case of light-through-walls experiments since a photon must convert to an axion and then convert back to a photon the rate goes as $g_{a\gamma\gamma}^4$.
Second, the operator in Eq.~\eqref{eqn: axion photon coupling} usually suppresses the signal in a possible experiment by the ratio of the size of the experiment over the wavelength of the axion (often squared).  This arises because $F \tilde{F}$ is a total derivative and therefore the operator in Eq.~\eqref{eqn: axion photon coupling} can be thought of as having a derivative on the axion field.  For high mass axions this is not a problem, microwave cavities can easily be the same size as the axion wavelength.  But for low mass axions this is a large suppression for any laboratory sized experiment.

\subsection{A New Operator for Axion Detection}

To detect low mass axions or ALPs we must avoid these problems.  We therefore propose using a different coupling instead of the one in Eq.~\eqref{eqn: axion photon coupling}.  
The QCD axion solves the strong CP problem, the problem that a nucleon electric dipole moment (EDM) would be generated by the $\theta$ parameter of QCD.  This parameter arises in the Standard Model lagrangian term $\theta G \tilde{G}$, where $G$ is the QCD field strength.  The QCD axion solves this problem essentially by turning $\theta$ into the dynamical axion field.
Thus the QCD axion is defined by its coupling $\propto \frac{a}{f_a} G \tilde{G}$.  This means the axion gives an effective $\theta$ angle.  This will then give rise to an EDM for nucleons sourced by the axion.  Because the axion is a dynamical field, this EDM will change in time, giving rise to unique signals.  We propose to search for this time-varying EDM as a new way to detect axions or ALPs.

This EDM can be expressed as the operator coupling the axion to nucleons $N$:
\begin{equation}
\label{eqn: axion EDM coupling}
\mathcal{L} \ni -\frac{i}{2} g_d \, a \, \bar{N} \sigma_{\mu \nu} \gamma_5 N F^{\mu \nu}.
\end{equation}
where $g_d$ is a coupling constant we introduce.  In general a new light particle or ALP could have a coupling like this as well.  The nucleon EDM generated by this operator is
\begin{equation}
\label{eqn: nucleon EDM}
d_n = g_d a
\end{equation}
where $a$ is the value of the local axion or ALP field at the position of the nucleus.  Thus, any nucleon in the axion or ALP dark matter will acquire an EDM proportional to the  dark matter field.  Note that the operator in Eq.~\eqref{eqn: axion EDM coupling} is a non-derivative coupling for the ALP so it naturally avoids the wavelength suppression discussed above for low mass axions or ALPs.


For the QCD axion the nucleon EDM created by its QCD coupling is determined in terms of the axion decay constant $f_a$:
\begin{equation}
\label{eqn: qcd axion nucleon EDM}
d_n^\text{QCD} \approx 2.4 \times 10^{-16} \frac{a}{f_a} e \cdot \cm
\end{equation}
with about a 40\% uncertainty \cite{Pospelov:1999ha}.  Thus, for the QCD axion our coupling $g_d$ is determined by the axion decay constant $f_a$ as 
\begin{equation}
\label{Eqn: QCD axion gd}
g_d^\text{QCD} \approx \frac{2.4 \times 10^{-16}}{f_a} \, e \cdot \cm \approx 5.9 \times 10^{-10} \left( \frac{m_a}{\text{eV}} \right) \GeV^{-2}
\end{equation}
where $m_a$ is the axion mass from Eq.~\eqref{eqn: axion mass}.  For an ALP of course, the coupling $g_d$ is in general arbitrary, independent of the mass of the ALP.

Note that one useful feature of searching for this EDM is that it is naturally a measurement of an amplitude (or phase), not a rate.  Thus the signal in an experiment will only be proportional to one power of $g_d$, or one power of $\frac{1}{f_a}$ for the QCD axion.  This is in contrast to all other experiments which are measuring rates and whose signals therefore are $\propto g_d^2$ or $\propto \frac{1}{f_a^2}$.  Measuring an amplitude and not a rate makes it much easier to push the sensitivity up to high $f_a$ (low axion couplings).

Further, the actual size of the EDM is set by the product $g_d a$, where $a$ is the local dark matter, axion or ALP, field.  As discussed in Section \ref{Sec:Overview} this is approximately $a \approx a_0 \cos \left( m_a t \right)$.  The amplitude of this field, $a_0$, is known if we require that this field makes up (all of) the local dark matter density
\begin{equation}
\label{eqn: dm abundance}
\rho_\text{DM} = \frac{1}{2} m_a^2 a_0^2 \approx 0.3 \frac{\GeV}{\cm^3}
\end{equation}
since the field $a$ is essentially a free scalar field with this mass term as the leading term in its potential.  This then determines the nucleon EDM generated by ALP dark matter from Eqs. \eqref{eqn: nucleon EDM} and \eqref{eqn: dm abundance} to be
\begin{equation}
\label{eqn: ALP EDM}
d_n = g_d \frac{\sqrt{2 \, \rho_\text{DM}}}{m_a} \cos \left( m_a t \right)  \approx \left( 1.4 \times 10^{-25} \, e \cdot \cm \right) \left( \frac{\text{eV} }{m_a} \right) \left( g_d \, \GeV^2 \right) \cos \left( m_a t \right)
\end{equation}

For the QCD axion, since the axion mass from Eq.~\eqref{eqn: axion mass} scales as $m_a \propto \frac{1}{f_a}$, taking the axion to be all of the dark matter fixes the effective $\theta$ angle of the axion to be independent of $f_a$:
\begin{equation}
\label{qcd axion theta}
\frac{a_0}{f_a} \approx 3.6 \times 10^{-19}
\end{equation}
So for the QCD axion dark matter, the nucleon EDM it induces from Eq.~\eqref{eqn: qcd axion nucleon EDM}, is actually independent of $f_a$
\begin{equation}
\label{eqn: qcd axion edm}
d_n^\text{QCD} \approx \left( 9 \times 10^{-35} \, e \cdot \cm \right)  \cos \left( m_a t \right).
\end{equation}
Thus we have found a physical effect that does not decouple as $f_a$ increases.  This useful fact allows experiments searching for this EDM to probe high $f_a$ axions.

Note that the EDM induced by the axion from Eq.~\eqref{eqn: qcd axion edm} is small.  Of course the EDM induced by a general ALP, Eq.~\eqref{eqn: ALP EDM}, is arbitrary.  For the QCD axion though, the EDM is about eight orders of magnitude smaller than the current bound on the static nucleon EDM, though these may improve in the future.  However the reason to believe that this EDM may be detectable is that it is not a static EDM.  The axion or ALP field oscillates with a frequency approximately equal to its mass Eq.~\eqref{eqn: axion mass}.  Thus the nucleon EDM from Eq.~\eqref{eqn: ALP EDM} also oscillates with this frequency.  This allows a resonant enhancement of the signal in an experiment (similar to ADMX) which can be many orders of magnitude.  Further, backgrounds for an oscillating signal are very different, and usually much more controllable, than backgrounds for a static signal.  The frequency of the oscillation is set by high energy physics, independent of anything in the laboratory setup.

Thus every nucleon bathed in axion (or this type of ALP) dark matter has an oscillating EDM with frequency set by the mass of the axion field.  Further, this oscillation will be in phase for all nucleons within the axion dark matter coherence length $\sim \frac{1}{m_a v}$, see Section \ref{Sec:Overview}.  One immediate idea is to search for the electromagnetic radiation given off by all these nuclear ``antennas."  We could not find a plausibly observable signal of this radiation, using either laboratory or astrophysical (e.g.~neutron star) sources.  One problem is that the radiation rate is proportional to the square of the EDM, and additionally to the small frequency or mass of the axion or ALP.

We have already proposed one idea using interferometry of cold molecules to detect axion dark matter using the nuclear EDM \cite{Graham:2011qk}.  We also believe that experiments based on nuclear magnetic resonance (NMR) techniques to measure spin precession may allow detection of ALPs and even the QCD axion over many orders of magnitude in parameter space \cite{NMR paper} (see Figure 2 of \cite{NMR paper} for sensitivity estimates).

\subsection{A New ALP Parameter Space}

\begin{figure}
\begin{center}
\includegraphics[width=6 in]{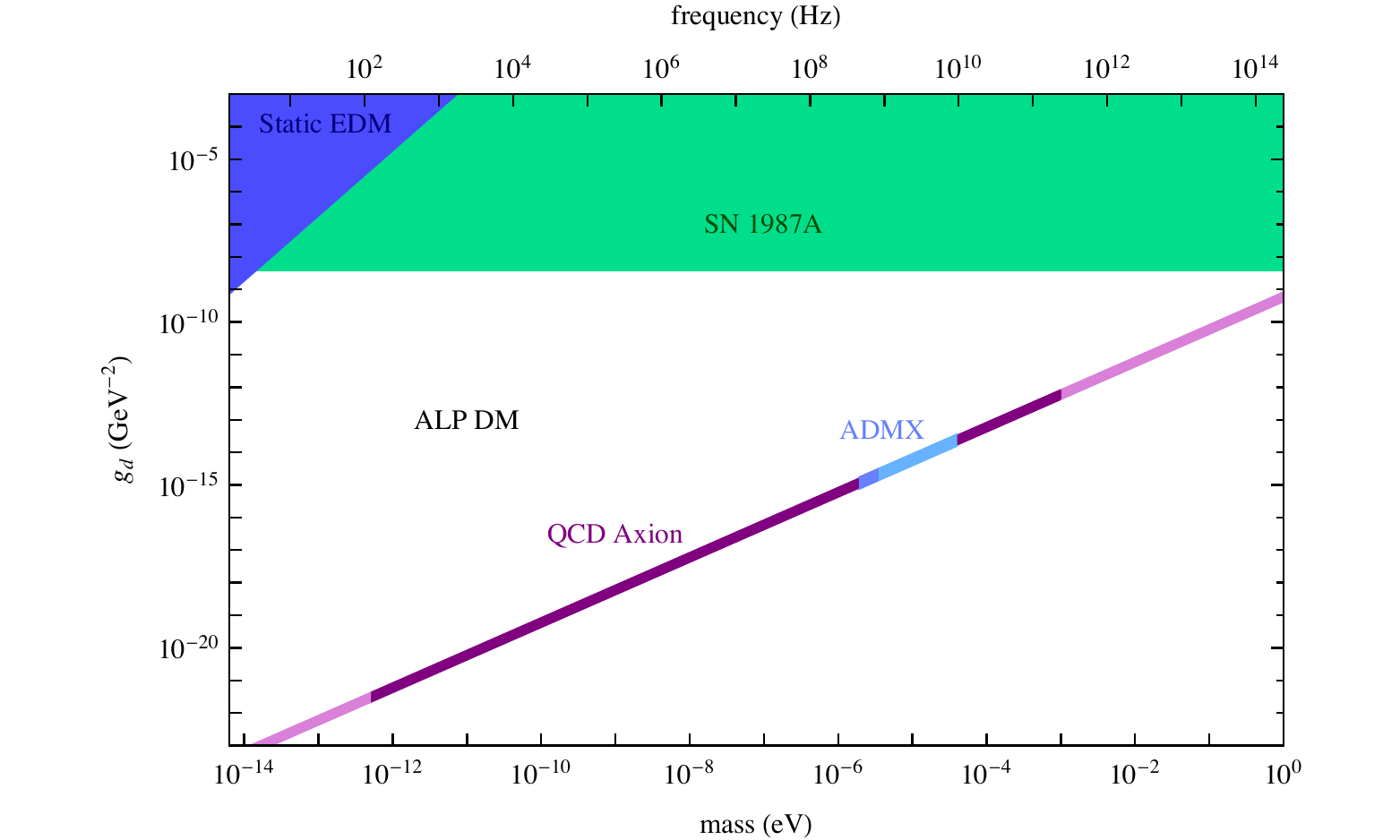}
\caption{ \label{Fig:EDM} ALP parameter space in EDM coupling Eq.~\eqref{eqn: axion EDM coupling} vs mass of ALP.  The green region is excluded by excess cooling in SN1987A.  The blue regions are excluded by the best static nucleon EDM experiments.  The purple band is the QCD axion region, with dark purple showing the most theoretically-motivated region for QCD axion dark matter.  The width of the band shows the uncertainty in the calculation of the axion-induced EDM and the axion mass.  The ADMX region shows the part of QCD axion parameter space which has been covered (darker blue) \cite{Asztalos:2009yp} or will be covered in the near future (lighter blue)  \cite{ADMXwebpage, snowdarktalk} by ADMX.  For the static EDM and ADMX bounds we assume that the ALP makes up all of the dark matter.  See also Figure 2 of \cite{NMR paper} for sensitivity of the proposed NMR experiment.}
\end{center}
\end{figure}

Using the EDM coupling in Eq.~\eqref{eqn: axion EDM coupling} naturally suggests a new parameter space in which to search for axion-like particles.
The EDM coupling is naturally generated for the QCD axion.  Of course, the scalar field $a$ in Eq.~\eqref{eqn: axion EDM coupling} does not have to be the QCD axion.  If not, it would be a type of ALP.  Most considerations of ALPs have focused on fields that couple to electromagnetism through the coupling in Eq.~\eqref{eqn: axion photon coupling}.  However there is no reason that an ALP has to couple only, or at all, through the electromagnetic coupling.  The space of ALPs that have EDM couplings as in Eq.~\eqref{eqn: axion EDM coupling} is a new parameter space that is worth exploring since this coupling motivates new experimental signatures and appears promising for axion detection.  Figure \ref{Fig:EDM} shows the parameter space for an ALP in the space of the EDM coupling $g_d$ in Eq.~\eqref{eqn: axion EDM coupling} versus the mass of the ALP.

The QCD axion may lie anywhere on the purple line in Figure \ref{Fig:EDM}.  It lies on a line because it has only one free parameter, $f_a$, which determines its mass and $g_d$ as in Eqs.~\eqref{eqn: axion mass} and \eqref{Eqn: QCD axion gd}.  The width of the QCD axion line shows an estimate of the theoretical uncertainty in the calculation of these quantities, as described above.  The darker purple section of the line shows where the QCD axion may make up the dark matter.  For values of $m_a \gtrsim$ meV the QCD axion cannot have enough abundance to make up all of the dark matter, as discussed in Section \ref{Sec:Overview}, though it may be a subdominant component.  The lower edge of the dark purple region, where $m_a \approx 5 \times 10^{-13}$ eV, is where the axion decay constant is around the Planck scale, $f_a = M_\text{pl} \approx 1.2 \times 10^{19}$ GeV.  It is not clear if it is possible to make models of the QCD axion with masses below this scale, since that would require $f_a > M_\text{pl}$.  For that reason we keep this area light purple.  It is certainly worth searching for such axions though, since although it is not obvious how to make a model of such an axion without a full understanding of quantum gravity, it may well be possible.

We show the ADMX constraints as the blue region in Figure \ref{Fig:EDM}.  Since ADMX searches for the axion electromagnetic coupling Eq.~\eqref{eqn: axion photon coupling} these constraints cannot properly be put on this figure for the QCD or EDM coupling.  But ADMX can constrain most QCD axion models, where the coupling to photons is related to the coupling to gluons.  So we show the ADMX constraints in Figure \ref{Fig:EDM} as constraints just on the QCD axion parameter space.  The darker blue represents the current bounds from ADMX \cite{Asztalos:2009yp} while the lighter blue represents the region ADMX will cover in the near future \cite{ADMXwebpage, snowdarktalk}.

The constraint from SN1987A arises from excess cooling caused by axion emission.  This constraint usually arises from the other axion-nucleon coupling Eqn.~\eqref{eqn:gaNN} as in \cite{Raffelt:2006cw}.  We adapted it to the higher dimension EDM operator Eqn.~\eqref{eqn: axion EDM coupling} by calculating the axion emission rate from the SN using the process $N + \gamma \to N + a$.  We approximate this cross section as $\sigma v \approx g_d^2 T^2$ where $T \approx 30$ MeV is the temperature of the SN.  As an approximation we assume the axion is produced with energy equal to the average photon energy in the SN $\approx \frac{\pi^4}{30 \zeta(3)} T \approx 2.7 \, T$.  The energy lose rate per unit volume is then $\approx 2.7 \, T n_\gamma n_N \sigma v$, where $n_N \approx \frac{1.8 \times 10^{38}}{\text{cm}^3}$ is the number density of nucleons and $n_\gamma \approx \frac{2 \zeta(3)}{\pi^2} T^3$ is the number density of photons in the supernova.  Dividing this by the mass density of the supernova $\rho \approx 3 \times 10^{14} \frac{\text{g}}{\cm^3}$ gives the cooling rate per unit mass.  This cooling rate is then compared to the bound $10^{19} \, \text{erg} \, \text{g}^{-1} \, \text{s}^{-1}$ from \cite{Raffelt:2006cw}.  This gives a bound on the EDM coupling $g_d \lesssim 4 \times 10^{-9} \, \GeV^{-2}$.  This bound is shown as the green region in Figure \ref{Fig:EDM}.  Of course this calculation is only a rough approximation, but it is good enough for our purposes since the parameter space we are interested in extends many orders of magnitude below this bound.  We leave a more precise calculation for future work.

The experiments searching for (static) nuclear EDMs have drastically reduced sensitivity to an oscillating EDM of the type we are considering.  These experiments gain sensitivity by integrating for relatively long periods of time compared to the period of the oscillating EDM we are considering, Eq.~\eqref{eqn: ALP EDM}.  Since this oscillating EDM has an average value of zero these static EDM experiments are not well-suited to searching for this signal of an ALP.  The limits set by the static EDM experiments are shown in the dark blue region of Figure \ref{Fig:EDM} \cite{Baker:2006ts, Harris:1999jx}.  These experiments have not done a specific search for an oscillating EDM so we calculated an approximate limit using their limits on the static nuclear EDM as follows.  We assume these experiments gain sensitivity linearly in time over their shot time.  They measure the total precession of the neutron spin when exposed to an electric field for a shot time, which is $t_\shot = 130$ s for \cite{Baker:2006ts}.  However since the oscillating EDM averages to zero, only the last fraction of a period of oscillation will cause a net precession.  Thus these experiments lose a factor of $\sim t_\shot m_a$ in sensitivity to an oscillating EDM.  Further, the signal of this oscillating EDM will be stochastic from shot to shot and so we assume roughly that the sensitivity will not improve with the number of shots.  We assume this means the static EDM experiments lose another factor of $\sim \sqrt{N_\shot}$ in their sensitivity.  The number of shots was $N_\shot = 221600$ in \cite{Baker:2006ts}.  So finally we estimate the limit these experiments set on the amplitude of an oscillating EDM is $\sim d_N^\static \left( t_\shot m_a \right) \sqrt{N_\shot}$ where $d_N^\static = 2.9 \times 10^{-26} \, \text{e} \cdot \cm$ is the limit they set on the static nucleon EDM.  Under the assumption that the ALP makes up all of the dark matter we can translate this to a bound on the coupling $g_d$ using Eq.~\eqref{eqn: ALP EDM}.  This gives a bound of $g_d \lesssim 1.8 \times 10^{19} \left( \frac{m_a}{\eV} \right)^2 \GeV^{-2}$, as shown in Figure \ref{Fig:EDM}.  This bound is only a rough approximation but it is good enough for our purpose since it is a very weak bound on these ALPs.  Ideally the experiments themselves would reanalyze their data to directly search for an oscillating EDM.  Since these experiments are looking for a static nucleon EDM they are simply not designed appropriately to look for the oscillating EDM signal we are discussing.

The EDM coupling in Eq.~\eqref{eqn: axion EDM coupling} is an interesting and useful one to consider for axion or ALP detection partially because it is a non-derivative coupling.  For the QCD axion this coupling arises completely naturally.  For an ALP this coupling may also be natural.  However it can induce a mass for the ALP through a two-loop diagram.  Thus the most natural part of such ALP parameter space, Figure \ref{Fig:EDM}, is either below the QCD axion line or above it and within several orders of it.  ALP parameter space that is many orders of magnitude above the QCD axion line may become tuned, though at some point of course is ruled out by SN1987A anyway.  
Of course, this concern applies to the scalar ALP model we have considered, other fields or models could change this.

Having ALP dark matter with the coupling we are considering changes the status of the nucleon EDM from a fundamental constant of nature to a parameter dependent on the local field value.  Thus we see that the nucleon EDM may be expected to change in time (and space), likely oscillating at high frequencies $\sim$ kHz to GHz.  It is thus important to consider the limits that existing experiments put on the parameter space in Figure  \ref{Fig:EDM}.  Further, this parameter space has not been considered before.  Therefore it is also important to design experiments which are optimized to search for this signal.  Beyond the cold molecule \cite{Graham:2011qk} and NMR techniques \cite{NMR paper} that we have considered there could be many possibilities for other experiments, for example using proton storage rings \cite{Semertzidis:2011qv, Orlov:2006su, Semertzidis:2003iq, yannis}.

\section{Axial Nuclear Moment}
\label{Sec: axial nuclear}
The third operator in \eqref{Eqn:Terms} gives rise to the coupling

\begin{equation}
\label{eqn:gaNN}
\mathcal{L} \supset g_\text{aNN} \left( \partial_\mu a \right) \bar{N} \gamma^\mu \gamma_5 N
\end{equation}
between the ALP and the axial nuclear current. For the QCD axion, this coupling usually exists to both protons and neutrons and is approximately $g_\text{aNN} \sim \frac{1}{f_a}$. Current bounds on this operator arise from two sources. First, this operator allows an accelerated nucleon to lose energy through ALP emission. These emissive processes are constrained by observations of the cooling rates of supernova, imposing an upper bound on $g_{aNN} \lessapprox 10^{-9} \text{ GeV}^{-1}$ \cite{Raffelt:2006cw}. Second, this operator leads to a force between nucleons through the exchange of ALPs. This force is spin dependent with a range $\sim m_a^{-1}$ \cite{Moody:1984ba}.  Such spin-spin interactions have been searched for using a variety of spin polarized targets, but the limits on $g_{aNN}$ from them \cite{Vasilakis:2008yn} are weaker than the constraints from supernova emission \cite{Raffelt:2006cw, Engel:1990zd} (see Fig. \ref{Fig:Nucleon}). 

The above effects do not require the presence of a background ALP field. In the presence of such a field, for example as ALP dark matter, the non-relativistic limit of this operator leads to the following term in the nucleon Hamiltonian
\begin{equation}
H_N \supset  g_{aNN} \vec{\nabla} a.\vec{\sigma_N}
\end{equation}
where $\sigma_N$ is the nucleon spin operator. Much like a spin precessing around a background magnetic field, this coupling  causes spin precession of a nucleon around the local direction of the ALP momentum $\vec{\nabla} a$.

The motion of the Earth through the galaxy leads to a relative velocity between it and the dark matter. As long as the nucleon spin is not aligned with this velocity, the spin will precess about this ALP dark matter ``wind". Since the magnitude of this relative velocity is the galactic virial velocity $v \sim 10^{-3} c$, the ALP has a momentum of $\vec{\nabla} a \sim 10^{-3} \, \partial_0 a $.  To leading order the ALP dark matter field is simply a free scalar field with low momentum which is oscillating in its potential so it is approximately $a \approx a_0 \cos \left( m_a t \right)$.  Thus $\partial_0 a$ has magnitude $a_ 0 m_a$ and oscillates with frequency $m_a$. The effective coupling in the nucleon Hamiltonian is
\begin{equation}
\label{eqn: nucleon precession rate from gann}
H_N \supset  g_{\text{aNN}}  \, m_a a_0 \cos \left(m_a t\right)  \, \vec{v} .\vec{\sigma_{N}} 
\end{equation}
The amplitude $a_0$ of the ALP field is constrained by the requirement that the energy density $\frac{1}{2} m_a^2 a_0^2$ in the ALP oscillations not exceed the local dark matter density $\rho_\text{DM} \sim  0.3 \, \frac{\GeV}{\cm^3}$. Hence, the maximum size of this perturbation is 
\begin{equation}
\label{eqn: numbers for nucleon precession rate from gann}
\Delta E \sim  g_\text{aNN} \, \sqrt{\rho_{\text{DM}}} \,  v \sim 3 \times 10^{-9} \text{ s}^{-1} \, \left(\frac{g_{aNN}}{ 10^{-9} \text{ GeV}^{-1}}\right) \left(\sqrt{\frac{\rho_{\text{DM}}}{0.3  \, \frac{\text{GeV}}{\text{cm}^3}}}\right)
\end{equation} 
oscillating at a frequency equal to the ALP mass $m_a \sim$ kHz - GHz. The expected coherence time for this oscillation is set by the ALP coherence time $\tau_a \sim \frac{1}{m_a v^2} \sim 1 \text{ s} \, \left(\frac{\text{MHz}}{m_a}\right)$, leading to a signal bandwidth $\sim 10^{-6} m_a$. 

\subsection{A Detection Strategy}
\label{subsec: det strategy}

\begin{figure}
\begin{center}
\includegraphics[width=3.5 in]{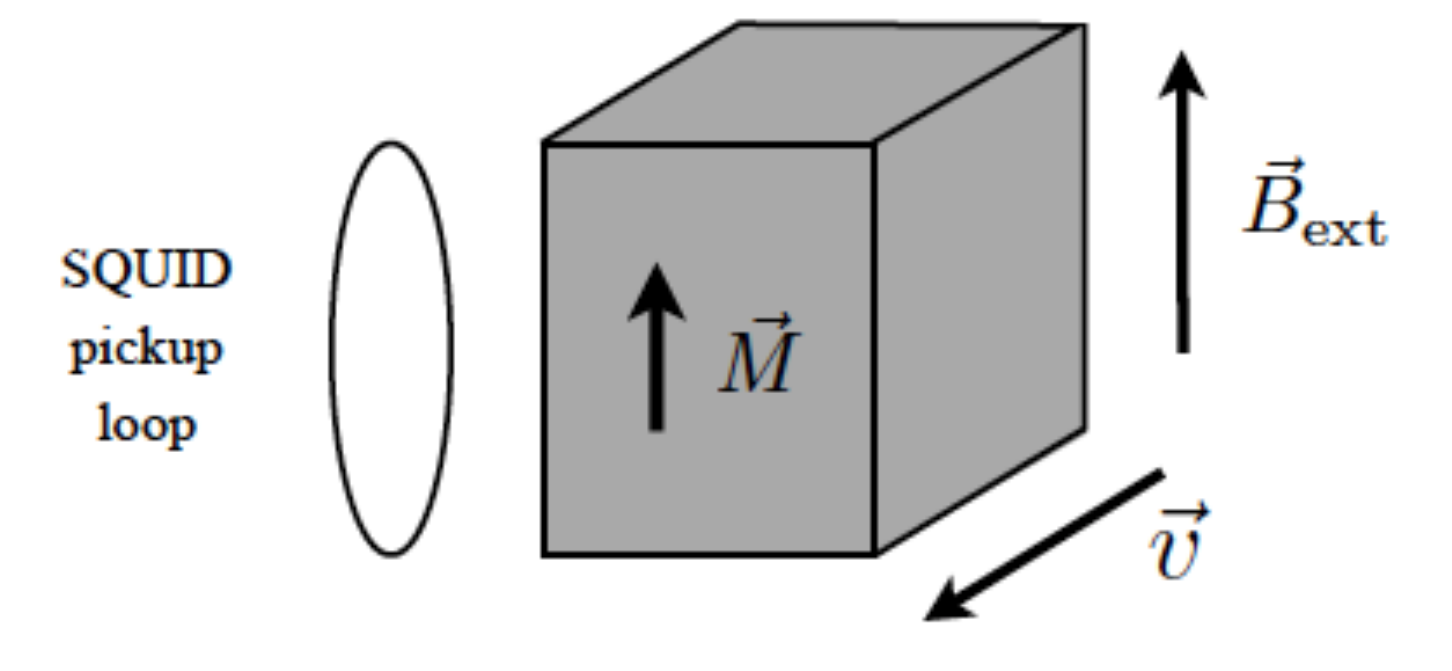}
\caption{ \label{Fig:setup} Geometry of the experiment, adapted from \cite{NMR paper}.  The applied magnetic field $\vec{B}_\text{ext}$ is collinear with the sample magnetization $\vec{M}$.  The relative velocity $\vec{v}$ between the sample and the dark matter ALP field is in any direction that is not collinear with $\vec{M}$.  The SQUID pickup loop is arranged to measure the transverse magnetization of the sample.}
\end{center}
\end{figure}

The detection of this small but time varying energy shift requires the development of new experimental techniques. While there may be many experimental avenues that could be pursued, we highlight the approach proposed in \cite{NMR paper} utilizing NMR techniques. In this approach, a sample of nuclear spin polarized material is placed with the polarization chosen along a direction that is not collinear  to the relative velocity $\vec{v}$ between the Earth and the dark matter, as in Fig.~\ref{Fig:setup}. An axial nuclear moment \eqref{eqn:gaNN} in the presence of a dark matter ALP field will cause the spins to precess around this relative velocity. This precession changes the magnetization of the material and can be measured  using precision magnetometers such as SQUIDs or SERFs. 

More specifically, the procedure is to polarize the nuclear spins of a sample of material in an external magnetic  ($\vBext$) to achieve a net magnetization. When this net magnetization is not collinear with the dark matter velocity, the spins will precess around this relative velocity.   Once they are no longer aligned with the external magnetic field,  they will precess around both the relative velocity and this magnetic field. Equivalently, the nuclear spins precess around the relative velocity as seen in a rotating frame in which the magnetic field is eliminated. This results (as seen in the lab frame) in a magnetization at an angle to the magnetic field, which precesses around this field with the Larmor frequency. This gives rise to a transverse magnetization, which can be measured with a magnetometer such as a superconducting quantum interference device (SQUID) with a pickup loop oriented as shown in Fig.~\ref{Fig:setup}. The transverse magnetization rotates at the Larmor frequency set by the external magnetic field.  When the ALP oscillation frequency is different from the Larmor frequency, no measurable transverse magnetization ensues. However, when the two frequencies coincide, there occurs a resonance akin to that in the usual NMR, where the spins precess around a transverse axis rotating at the Larmor frequency \cite{BudkerBook}. This effect enhances the precessing transverse magnetization that can be detected with the SQUID magnetometer.
The magnitude of the external magnetic field ($\Bext$) is swept to search for a resonance.  At time $t=0$ the spins are prepared along $\vBext$, then the magnitude of the transverse magnetization is given by
\begin{equation}
\label{eqn: magnetization signal}
M(t) \approx n p \mu \, \left(g_{\text{aNN}} \sqrt{2 \, \rho_{\text{DM}}} v\right) \, \frac{\sin \left( \left( 2 \mu \Bext - m_a \right) t \right)}{2 \mu \Bext - m_a}  \sin \left(  2 \mu \Bext t \right),
\end{equation}
where $n$ is the number density of nuclear spins, $p$ is the polarization, and $\mu$ is the nuclear magnetic dipole moment.  The resonant enhancement occurs when $2 \mu \Bext \approx m_a$. Taking $n \sim \frac{10^{22}}{\text{cm}^3}$, $p \sim \mathcal{O}\left(1\right)$,  $\mu \sim \mu_N$ (the nuclear Bohr magneton) and interrogation time $ t \sim \tau_a \sim \frac{10^{6}}{m_a}$ (the ALP coherence time), we get
\begin{equation}
M \approx 2 \times 10^{-14} \, \text{T} \, \left(\frac{g_{\text{aNN}}}{10^{-10} \, \text{GeV}^{-1}}\right) \, \left(\frac{\text{MHz}}{m_a}\right)
\end{equation}
This magnetic field is above the sensitivity of modern SQUID and atomic SERF magnetometers that typically have sensitivities $\sim 10^{-16} \frac{\text{T}}{\sqrt{\text{Hz}}}$. 

The axial nuclear moment oscillates at a frequency set by particle physics, independent of the experimental setup.  This distinguishes the signal from many possible backgrounds.  For example,  control over noise sources is only required over the signal's relatively high frequency range (kHz - MHz) and narrow bandwidth ($\sim 10^{-6} \, m_a$). Further, though the induced axial nuclear moment is small, its oscillation at laboratory frequencies enables resonant schemes that boost the signal significantly. 

This idea is based on and very similar to the one proposed in \cite{NMR paper} to detect the time varying EDM induced by the dark matter axion. However, in this case, since we are not searching for an EDM, we do not need a material with a large Schiff moment nor do we need to expose it to significant electric fields. The techniques described in \cite{NMR paper} to achieve large nuclear polarizations and quality factors for the NMR resonance can also be employed in this case. Further, noise sources such as the intrinsic magnetization noise of the sample and strategies to mitigate them should also be similar to the discussions of \cite{NMR paper}. 

\begin{figure}
\begin{center}
\includegraphics[width=6 in]{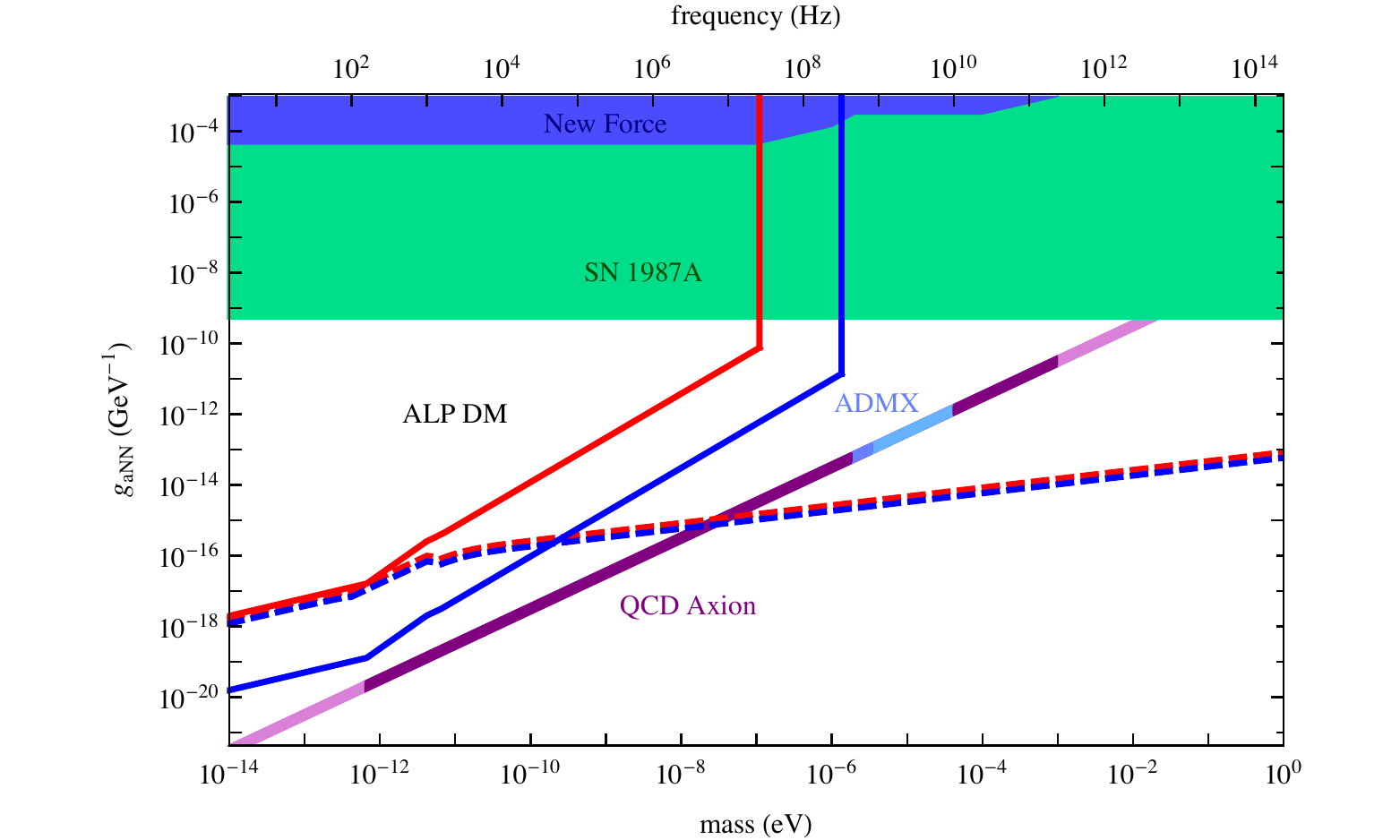}
\caption{ \label{Fig:Nucleon} ALP parameter space in pseudoscalar coupling of axion to nucleons Eqn.~\eqref{eqn:gaNN} vs mass of ALP.  The purple line is the region in which the QCD axion may lie.  The width of the purple band gives an approximation to the axion model-dependence in this coupling.  The darker purple portion of the line shows the region in which the QCD axion could be all of the dark matter and have $f_a < M_\text{pl}$ as in Figure \ref{Fig:EDM}.  The green region is excluded by SN1987A from \cite{Raffelt:2006cw}. The blue region is excluded by searches for new spin dependent forces between nuclei \cite{Vasilakis:2008yn}. The red line is the preliminary sensitivity of an NMR style experiment using Xe, the blue line is the sensitivity using $^3\text{He}$.  The dashed lines show the limit from magnetization noise for each sample.  These lines assume the parameters in Table \ref{Tab: experiments}. The ADMX region shows the part of QCD axion parameter space which has been covered (darker blue) \cite{Asztalos:2009yp}  or will be covered in the near future (lighter blue) \cite{ADMXwebpage, snowdarktalk} by ADMX.}
\end{center}
\end{figure}

Figure \ref{Fig:Nucleon} shows constraints on $g_\text{aNN}$ and the potential sensitivity of our proposals. The width of the line shows axion model-dependence in the axion-nucleon coupling.  The solid lines are preliminary sensitivity curves with the sensitivity limited by magnetometer noise. Both lines assume samples of volume $\left(10 \text{ cm}\right)^3$ with 100 percent nuclear polarization. Other sample parameters are described in Table \ref{Tab: experiments}.  The dashed lines show the limits from sample magnetization noise, so where they are higher than the corresponding solid line, they are the limit on sensitivity.  The solid curves are cutoff at high frequencies by the requirement that the Larmor frequency be achievable with the assumed maximum magnetic field.

\begin{table}
\renewcommand{\arraystretch}{1.1}
\begin{center}
\begin{tabular}{|l|c|c|c|c|c|c|}
\hline
& Element & Density  & Magnetic Moment  & $T_2$ & Max. B & Magnetometer   \\
& & ($n$) & ($\mu$) & & & Sensitivity \\
\hline
1. & Xe & $1.3 \times 10^{22} \frac{1}{\cm^{3}}$ & $0.35 \, \mu_N$ & 100 \text{s} & 10 T & $10^{-16} \frac{\text{T}}{\sqrt{\text{Hz}}}$ \\
2. & $^3$He & $2.8 \times 10^{22} \frac{1}{\cm^{3}}$ & $2.12 \, \mu_N$ & 100 \text{s} & 20 T & $10^{-17} \frac{\text{T}}{\sqrt{\text{Hz}}}$\\
\hline
\end{tabular}
\caption{\label{Tab: experiments} The parameters used for the sensitivity curves shown in Figure \ref{Fig:Nucleon}.  The first row corresponds to the upper (red) lines in the figure while the second row is the lower (blue) lines.  For the Xe experiment we used the average magnetic moment from the naturally occurring abundances of $^{129}$Xe and $^{131}$Xe.  The sixth column shows the maximum magnetic field that is assumed, which is relevant only for setting the upper frequency limit on the curves.  The last column shows the assumed magnetometer sensitivity.}
\end{center}
\end{table}

Note that there are many ways to verify a positive signal in such an experiment, the same ways as described in \cite{NMR paper}.  If a positive signal is found, the scan can be stopped and that particular frequency can be explored for much more time than was needed to observe it in the scanning mode.  Thus one can effectively make many measurements of the dark matter signal.  This has several interesting consequences.  In particular, the signal from operator of Eqn.~\eqref{eqn:gaNN} is proportional to the spatial derivative of the axion field, i.e.~the local axion velocity.  This is unlike the case for the EDM operator, Eqn.~\eqref{eqn: axion EDM coupling} (our NMR proposal of \cite{NMR paper}) or the photon coupling Eqn.~\eqref{eqn: axion photon coupling} (used in ADMX).  Hence, if the ALP signal can be observed through this operator, Eqn.~\eqref{eqn:gaNN}, we will actually have a directional dark matter detector.  One could observe simultaneously with 3 different samples with perpendicular magnetization directions (or just vary the magnetization direction using one sample).  This would give us the local axion velocity.  Within the axion coherence length, the wavelength $\sim \frac{1}{m_a v}$, all experiments must agree on this measured direction of the axion velocity.  So this is another check on a positive signal.  But it also gives much more information since it tells us about the velocity structure of the dark matter.  At any one instant of time the local velocity may appear random and changes on a timescale of order the axion coherence time $\tau_a \sim \frac{1}{m_a v^2}$.  However, if the signal is folded on a yearly period or a daily period, the average velocity should modulate exactly with the Earth's velocity around the sun or rotational velocity around its axis respectively.  This would be yet another check that the signal is correct.  Even using the EDM coupling or the photon coupling could lead to interesting knowledge about the dark matter velocity profile including knowledge of local streams, as has been pointed out for ADMX \cite{Sikivie:1992bk, Duffy:2006aa, Hoskins:2011iv}, because of the high frequency resolution.  However when using the pseudoscalar nucleon coupling, Eqn.~\eqref{eqn:gaNN}, we have something more, we have a directional detector so we learn information about the full velocity distribution of the dark matter.

It is very interesting that even this experiment can get close to the QCD axion over a very large range of axion masses and further can cover a large piece of ALP parameter space.  Also, very importantly, the fundamental limit from magnetization noise can be reduced by using samples with larger volumes \cite{NMR paper}. This scheme could thus potentially allow detection of the QCD axion over an interesting range of higher masses through the use of improved magnetometers. 

This NMR technique appears to have the capability to probe hitherto unconstrained ALP dark matter parameter space when the ALP couples to nuclear moments such as the electric dipole moment \cite{NMR paper} or the axial nuclear moment. While constraints from current laboratory experiments for these ALP induced nuclear moments are much weaker than astrophysical limits, this search for  ALP dark matter probes regions well beyond these limits (see Fig. \ref{Fig:Nucleon}).

\section{Axial Electron Moment}
\label{Sec: axial electron}

Much like the axial nuclear moment discussed above, ALPs can also couple to electrons through the third operator in \eqref{Eqn:Terms} giving rise to the interaction 
\begin{equation}
\label{eqn:gaee}
\mathcal{L} \supset  g_\text{aee} \, \partial_{\mu}a \left( \bar{e} \gamma_5 \gamma^{\mu} e \right).
\end{equation}
This coupling is very similar to the nucleon coupling in Eqn~\eqref{eqn:gaNN} and leads to similar effects. The QCD axion generally has this coupling with $g_{aee} \sim \frac{1}{f_a}$, though it can be fine-tuned to zero.  Astrophysics constrains  $g_{aee} \lessapprox 10^{-10} \text{ GeV}^{-1}$  from bounds on the cooling of white dwarves \cite{Raffelt:2006cw}. This interaction also gives rise to spin dependent dipole - dipole forces between electrons. However, bounds from such searches are significantly weaker than the astrophysical limits on this coupling \cite{Dobrescu:2006au, electronspin}. 

\begin{figure}
\begin{center}
\includegraphics[width=6 in]{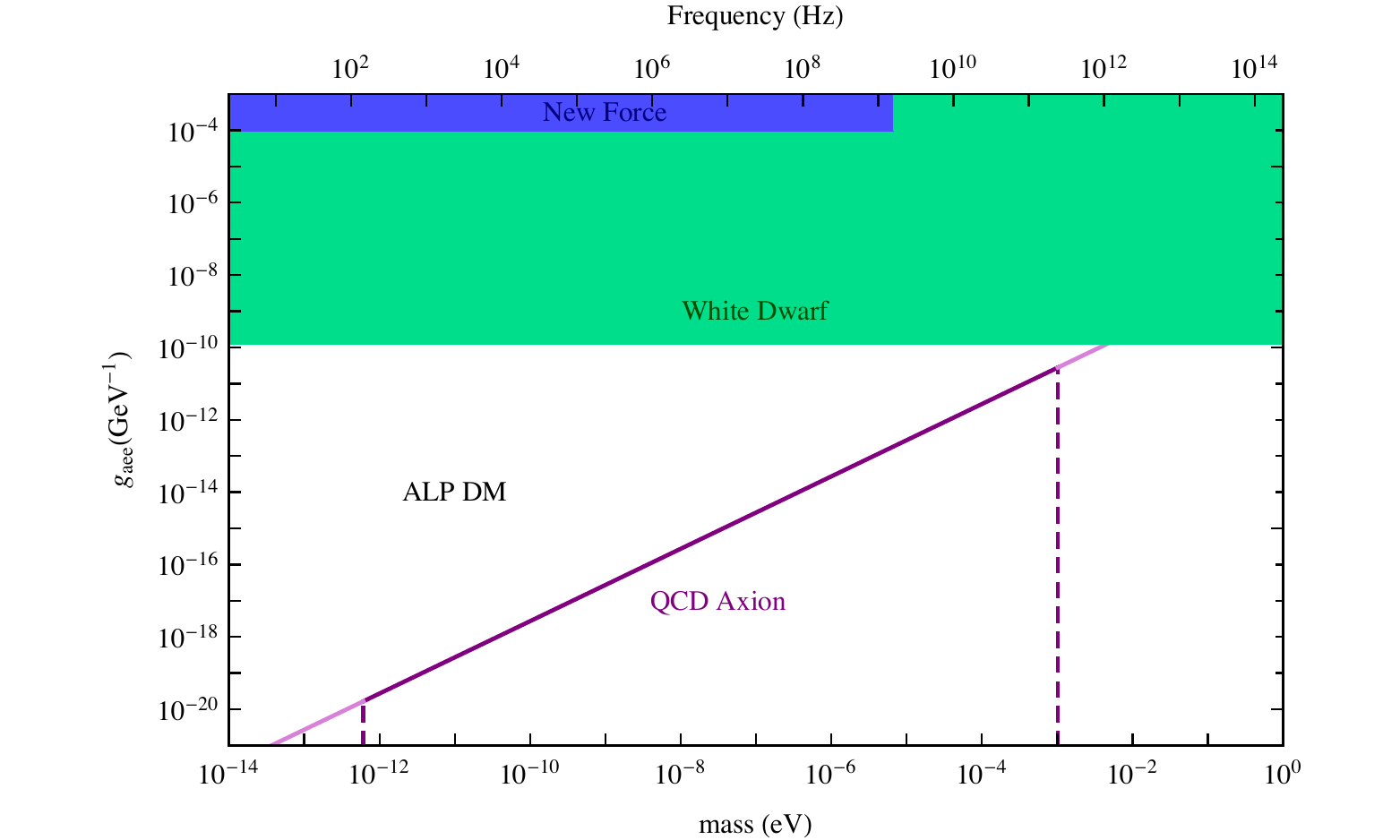}
\caption{ \label{Fig:Electron} ALP parameter space in pseudoscalar coupling of axion to electrons Eqn.~\eqref{eqn:gaee} vs mass of ALP.  The green region is excluded by White Dwarf cooling rates from \cite{Raffelt:2006cw}. The blue region is excluded by searches for new spin dependent forces between electrons \cite{Dobrescu:2006au, electronspin}. The region below the solid purple line shows the possible parameter space for a QCD axion, with the region bounded by darker purple lines being the region where the QCD axion could be all of dark matter and have $f_a < M_\text{pl}$. The frequency range of the QCD axion covered by ADMX is identical to the range plotted in Figure \ref{Fig:Nucleon}. }
\end{center}
\end{figure}

Similar to the axial nuclear moment, in the presence of a background dark matter ALP field, the non-relativistic limit of this operator leads to the following term in the electron Hamiltonian 
\begin{equation}
H_e \supset  g_{aee} \vec{\nabla} a.\vec{\sigma_e}
\end{equation}
where $\sigma_e$ is the electron spin operator. An electron spin that is not aligned with the ALP dark matter ``wind" will then precess due to the coupling 
\begin{equation}
\label{eqn: electron precession rate from gaee}
H_e \supset g_\text{aee}  \, m_a \, a_0  \cos \left(m_a t\right) \, \vec{v}.\sigma_{e}
\end{equation}
Using the constraint that the energy density in the ALP oscillations not exceed the local dark matter density, this perturbation is of size 
\begin{equation}
\label{eqn: numbers for electron precession rate from gaee}
\Delta E \sim g_\text{aee} \, | \vec{\nabla} a | \sim g_\text{aee} \, v \, \sqrt{\rho_{\text{DM}}} \sim 3 \times 10^{-9} \text{ s}^{-1} \, \left(\frac{g_{aee}}{ 10^{-9} \text{ GeV}^{-1}}\right) \left(\sqrt{\frac{\rho_{\text{DM}}}{0.3  \, \frac{\text{GeV}}{\text{cm}^3}}}\right)
\end{equation} 
This perturbation also oscillates at a frequency equal to the ALP mass $m_a \sim$ kHz - GHz, with an expected bandwidth $\sim 10^{-6} m_a$.

We have not been able to invent techniques that could  probe this unconstrained parameter space of ALP dark matter. We show constraints on this coupling in Figure \ref{Fig:Electron}. The solid purple line in the figure shows the largest value that $g_{\text{aee}}$ could take for the QCD axion.  Since $g_{\text{aee}}$ is model dependent, it could in principle be tuned to zero, though it is generally expected to be close to the purple line.  As in Figures \ref{Fig:EDM} and \ref{Fig:Nucleon} the darker purple portion shows the part of QCD axion parameter space where the axion may be all of the dark matter and has $f_a < M_\text{pl}$.  In this figure this region is bounded by the solid dark purple on top and the dashed lines on the sides. For a general ALP, there is no such expectation and the coupling could lie anywhere on the unconstrained portion of Figure \ref{Fig:Electron}. Experimental techniques to probe time varying electron axial moments could thus probe an unexplored range of ALP dark matter.

\section{Conclusions}
\label{Sec: conclusions}

All previous axion detection experiments have been based on the axion-photon coupling in Equation \eqref{eqn: axion photon coupling}.  We have considered several new operators for axion and ALP detection in Equations \eqref{eqn: axion EDM coupling}, \eqref{eqn:gaNN}, and \eqref{eqn:gaee} in Sections \ref{Sec: EDM}, \ref{Sec: axial nuclear}, and \ref{Sec: axial electron}.  For the QCD axion the EDM operator arises from the axion-gluon coupling $\propto \frac{a}{f_a} G \tilde{G}$.  We mapped out the parameter spaces for these operators including finding the current constraints in Figures \ref{Fig:EDM}, \ref{Fig:Nucleon}, and \ref{Fig:Electron}.  These operators suggest new ways to search for axion and ALP dark matter.  For the EDM coupling we previously proposed an experiment using cold molecules \cite{Graham:2011qk}.  These operators suggest promising detection strategies using spin precession, NMR-based, techniques which we discuss in detail in \cite{NMR paper}.

For the QCD axion, high-scale decay constants $f_a$, or masses below $\sim \mu \eV$, make up a well-motivated part of parameter space but are very challenging to detect with current experiments.  Use of these new operators may allow detection of QCD axion dark matter over a wider range of its parameter space, especially for $f_a$ near the fundamental GUT or Planck scales.  In particular the EDM operator Eq.~\eqref{eqn: axion EDM coupling} may be the most promising.  Because it is a non-derivative operator, it avoids the axion wavelength suppressions that plague the use of any other axion coupling for detecting low mass axions.

We have argued that it is useful to think of ALP dark matter produced through the misalignment mechanism as a classical field with an oscillating vacuum expectation value (VEV).  The interaction of a single axion or ALP particle with a detector may be too weak to observe.  But thinking of the ALP as a background field motivates searching for the coherent effects of the interaction of the entire classical scalar field with the detector.  For example, as we have shown, the ALP field may cause an oscillating nucleon EDM proportional to the classical VEV of the field, a collective effect of all the ALP `particles' comprising the field.  Or the ALP field may induce axial moments for nucleons or electrons, causing their spins to precess around the gradient of  the field.  

The continuous, coherent nature of these effects also enable secondary tests that can confirm the ALP dark matter origin of a signal in such experiments. As pointed out in \cite{NMR paper}, a signal in one sample can be correlated with another that is within the de-Broglie wavelength ($\gg 100$ m) of the ALP field. Further,  a positive signal in a particular bin can be verified by tuning the experiment to that bin and spending additional time to observe the build up of the signal in that bin. Since the assumed scanning time at any particular frequency bin is rather short ($\sim 10$ s), additional time can be spent in some bins without significant loss of efficiency.  As discussed earlier in Section \ref{Sec: axial nuclear}, the spin dependent nature of the ALP coupling to axial currents can be exploited to detect the direction of the dark matter wind for such ALPs. These effects are very different from the single, hard, particle scatterings which are used to search for WIMP dark matter.  For WIMP direct detection the signal is a stochastic energy deposition event in the detector.  The effects we propose searching for are not dominantly energy-deposition signals.  They are the continuous, coherent effects of the entire ALP field on the sample.


We considered ways to search for axion or ALP dark matter.  Similarly to ADMX, such signals benefit from requiring only one insertion of the small coupling between the axion or ALP and the Standard Model fields.  These couplings, $g_{a\gamma\gamma}$, $g_d$, $g_\text{aNN}$, and $g_\text{aee}$ in Equations \eqref{eqn: axion photon coupling}, \eqref{eqn: axion EDM coupling}, \eqref{eqn:gaNN}, and \eqref{eqn:gaee} respectively, are exceedingly small.  For the QCD axion they are all $\propto f_a^{-1}$, where $f_a$ is a high scale.  By contrast, light-through-walls and spin-dependent force experiments require two insertions of these couplings.  In such experiments the axion or ALP must be sourced (either by the laser or the source mass) and then must interact again to be detected.  The Feynman diagram would have two insertions of this operator and so the amplitude for the process is suppressed by the relevant coupling squared.  The light-through-walls experiments measure a rate and so are suppressed by the coupling to the fourth power.  This is why experiments searching for axion or ALP dark matter such as ADMX or through the effects we propose are sensitive to significantly smaller couplings and may even reach the QCD axion.

We have proposed a new type of dark matter signal to search for: the rapid oscillation of some parameter, e.g. the nucleon EDM.  In general, such a signal may arise from any type of modulus dark matter.  The QCD axion provides a well-motivated example of such a modulus, but many others are possible.  For example perhaps the dark matter is a modulus of electric charge, in which case the fine-structure constant would oscillate in time.  Unlike the current experiments searching for time-variation of $\alpha$ which look on timescales of years or more, the most motivated variation is on much faster timescales, frequencies of $\sim$ kHz to GHz or more.  Further, for definiteness here we have considered scalar fields.  However it is also possible that other fields (e.g.~vectors) may provide a natural realization of the experimental signatures we have considered.

Although we have considered some experimental designs to detect these signals of axion and ALP dark matter, it seems likely that many other experiments are also possible.  For example, some static EDM experiments may be modifiable to search for oscillating EDMs.  It would be valuable to make progress covering the ALP parameter spaces of Figures \ref{Fig:EDM}, \ref{Fig:Nucleon}, and \ref{Fig:Electron} and reaching towards the QCD axion.  Once we start considering the parameter space for this new type of signal, it becomes clear that there is a large, new class of dark matter direct detection experiments that have not been considered before.


Over the last couple decades, WIMP direct detection experiments have made tremendous progress, improving sensitivities by many orders of magnitude.  A similar improvement in the search for axion dark matter may be possible with new experiments designed to search for the coherent field effects we have described.  The axion is an excellent dark matter candidate.  Hopefully consideration of these types of signals will open new avenues to its discovery.

\section*{Acknowledgments}
We would like to thank Dmitry Budker, Hooman Davoudiasl, Savas Dimopoulos, Micah Ledbetter, Yannis Semertzidis, Alex Sushkov, and Scott Thomas for useful discussions.  SR was supported by ERC grant BSMOXFORD no. 228169.

\end{document}